\documentclass[english,aps,prl,twocolumn]{revtex4}
\usepackage{amssymb}

\makeatletter

\usepackage{babel}
\makeatother

\begin{document}

\title{ A new non-perturbative approach to Quantum Brownian Motion}

\author{Subhasis Sinha, P. A. Sreeram}
\affiliation{Indian Institute of Science Education and Research, HC Block, Sector III, Salt Lake, Kolkata-700106, India}
\begin{abstract}
Starting from the Caldeira-Leggett (CL) model \cite{cl}, we derive the equation describing the Quantum Brownian motion, which has been originally proposed by Dekker \cite{dekker} purely from phenomenological basis containing extra anomalous diffusion terms. 
Explicit analytical expressions for the temperature dependence of the diffusion constants are derived.
At high temperatures, additional momentum diffusion terms are suppressed and 
classical Langivin equation can be 
recovered and at the same time positivity of the density matrix(DM) is satisfied. At low temperatures, the diffusion constants have a finite positive value, however, below a certain critical temperature, the Master Equation(ME) does not satisfy the positivity condition as proposed by Dekker.
\end{abstract}
\maketitle

The problem of Quantum Brownian Motion is a long-standing and challenging problem \cite{ingold-chaos,ford1},
as it forms the underlying basis of non-equilibrium phenomena such as 
dissipative and relaxation 
dynamics of quantum systems \cite{sdg,weiss}. Quantum dissipative dynamics 
has application in a wide 
variety of problems, starting from quantum cosmological models 
\cite{halliwell, zurek} to reaction rate theory \cite{hanggi-rrt}. Unlike the classical nonequilibrium problem, quantum dynamics has additional complexity 
due to the Heisenberg Uncertainty Principle. Numerous attempts have been 
made to address this problem, in a variety of ways. These include 
semiclassical approaches using Wigner distribution function \cite{debshankar}, phenomenological models \cite{dekker,gao,diosi},
and using Boltzman's collision terms \cite{vacchini}. On the other hand, 
there have been attempts to obtain the Master Equation (ME) describing the 
time evolution of DM of open quantum systems through toy microscopic models
\cite{cl,spinboson,zurek}, where the environment is modelled by 
noninteracting particles or fields in thermal equilibrium. The ME is obtained 
for the reduced density matrix(DM) of the system by integrating out the 
environment degrees of freedom. New form of ME's are obtained from microscopic models where the diffusion constants have explicit time dependence 
\cite{zurek,paz}.
Independently, a class of ME's have been put forward, \cite{lindblad} purely from mathematical consideration, which 
guarantee the positivity of the density matrix. In other words, the time 
evolution of the system takes place only through the physical states, and the ME is commonly known to belong to the Lindblad class. Most of the MEs 
describing the non-equilibrium quantum dynamics suffers from the fact that 
either they do not belong to Lindblad class or correct classical limit can not be recovered at high temperatures.
The ME derived in the original paper by CL\cite{cl}, does 
not satisfy the positivity condition at high temperatures, although it 
recovers the classical Langevin dynamics at high temperatures.

Independently a phenomenological model of quantum Brownian dynamics 
has been proposed by Dekker \cite{dekker}, which contains extra terms 
describing momentum diffusion apart from ordinary diffusion, and the 
diffusion constants have to satisfy some conditions in order to preserve 
the Heisenberg's uncertainty condition as well as 
positivity of the DM \cite{positivity1,positivity2}. However Dekker's model 
of quantum 
Brownian motion is 
purely phenomenological and the microscopic origin of the extra momentum 
diffusion constants are not known. Moreover, most of the derivations of the ME's 
rely on the fact that the autocorrelation function of the random force on 
the Brownian particle is short ranged at high temperature and that the system 
is assumed to be Markovian. But in practice, the effective action obtained from 
CL model has a memory kernel which decays as a power law at low temperatures \cite{leggett2}. 
In fact, the random force in Quantum Brownian Motion is an operator and the force-force
autocorrelation function has been evaluated exactly, which shows clear deviation from the 
Markovian limit \cite{ford2}.
This nonlocality makes the problem very complex. 
In this letter we take a new approach, to analytically derive Dekker's  form 
of quantum Brownian motion, starting from the CL model, where the analytical 
expressions of the diffusion constants are obtained for all temperatures. 
We show that at high temperatures, the ME satisfies the positivity condition 
and at the same time reproduces the classical Langevin dynamics. Although, 
the diffusion constants are finite and well behaved at all temperatures, the 
positivity of DM breaks down below a certain temperature depending on the 
damping rate of the system. This is not very unexpected, since we know that 
at low temperatures, the long range memory effect becomes very important. 
This form of quantum Brownian motion is applicable to describe the relaxation phenomena and nonequilibrium evolution of quantum systems, even at 
sufficiently low temperatures.
We propose a new method to tackle the problem of nonlocality, using the 
equation of motion of the canonical coordinates. In this letter, we mainly 
focus on the quantum brownian dynamics in free space, however, our technique 
can, in principle, be extended to confined quantum particles.


A very well known description for the dissipative phenomena and relaxation
dynamics of classical system is given by the Langevin equation : 
\begin{equation}
\ddot{q}(t)=-\gamma\dot{q}(t)+\theta(t)
\label{Langevin}
\end{equation}
where, $q(t)$ is the position of the Brownian particle,
the dot denotes derivative with respect to time, $\gamma$ is the
damping constant and $\theta(t)$ is related to the fluctuating
force acting on the particle, whose autocorrelation function is given
by, \begin{equation}
\langle\theta(t)\theta(t^{\prime})\rangle=\Gamma\delta(t-t^{\prime})
\label{autocorrelation}
\end{equation}
where $\Gamma$ and $\gamma$ are related by the Fluctuation-Dissipation
Theorem, which states that $\Gamma$ = $2k_{B}$ T M$\gamma$, where
T is the temperature of the bath and M is the mass of the Brownian
particle. However, in most cases, the microscopic details behind this
dissipation are not well known.

The simplest microscopic model which describes the dissipative motion
of a heavy particle in presence of a heat bath was first put forward
by Ford, Kac and Mazur \cite{fkm} and later on applied to quantum
systems by Caldeira and Legget \cite{cl}.
In this model, full Hamiltonian has three different parts, system Hamiltonian
$ H_{A}$, the Hamiltonian of the heat bath $H_{B}$ and the Hamiltonian $H_{I}$
describing the linear coupling between the system and the bath, as given below
\begin{eqnarray}
H_{A} & = & \frac{p^{2}}{2M}+V(q)\\
H_{B} & = & \sum_{i}\left(\frac{P_{i}^{2}}{2m_{i}}+\frac{1}{2}m_{i}\omega_{i}^{2}Q_{i}^{2}\right)\\
H_{I} & = & q\sum_{i}C_{i}Q_{i}\end{eqnarray}

where, $q$ and $p$ are the position and momentum operators of the
heavy particle having mass $M$, $V(q)$ is the potential in which
the heavy particle is moving, $Q_{i}$ and $P_{i}$ are the position
and momentum operators of the bath oscillators whose mass and frequency
of oscillation are given by $m_{i}$ and $\omega_{i}$ respectively
and $C_{i}$ is the coupling strength between the heavy mass and the
$i^{th}$ oscillator in the bath. 
%
Following the work of Ford et. al., the time evolution of the position
operator can be written as, 
\begin{equation}
M\ddot{q}+\int_{0}^{t}dt^{\prime}\alpha(t-t^{\prime})\dot{q}(t^{\prime})+
V^{\prime}(q)=F(t)
\label{quant-lang}
\end{equation}
where the prime denotes the derivative with respect to the position
variable. The operator valued random force, $F(t)$ is related to
the statistical distribution of the bath variables $Q_{i}$ and $P_{i}$s.
%
In order to recover the classical Brownian dynamics with short time 
memory effect at high temperatures,
the frequency dependence of the coupling is typically taken to be
$C_{i}^{2}(\omega)/(m_{i}\omega_{i}^{2})$ $\propto$ $M\gamma$,
where $\gamma$ is the damping constant, and the summation is then
replaced by an integral over $\omega$. According the Ehrenfest theorem,
if the position and momentum operators are replaced by their classical
equivalents, then Eq. \ref{quant-lang} translates to the classical
Langevin Equation in the limit of $\hbar$ $\rightarrow$ 0, where
the force autocorrelation function takes the form of Eq. \ref{autocorrelation}
with $\Gamma = 2 M \gamma k_{B} T$\cite{sdg} obeying classical fluctuation-dissipation
theorem. 

The
essential difference between the classical Langevin equation and the
Quantum Master Equation is the fact that in the classical case, the
fluctuations in canonical coordinates are controlled only by the scales
set by the temperature and the dissipation constant. Whereas, in the
Quantum case, there is an additional scale given by $\hbar$ which
appears due the Heisenberg Uncertainty
Principle. Moreover, the
evolution of the operators themselves have to be unitary. Apart from the
semiclassical techniques, new
methods have been proposed to take into account the constraints imposed by
'uncertainty principle' in
the quantum case\cite{senitzky}. 
Among these methods, ME of the density matrix $\rho$ is more suitable to 
describe dissipative dynamics of quantum systems. The quantum dynamics which is governed by the
time dependent Schr\"{o}dinger equation describes the pure state, whereas
the dissipative dynamics introduces the concept mixed state, where
trace of $\rho^{2}$ is less than unity. The evolution equation of
the reduced density matrix of the system can be obtained by tracing
out the bath degrees of freedom. 

Using the Feynman-Vernon method \cite{feynman}, the dissipative dynamics of the reduced DM
can be written in terms of the influence functional, which can be obtained after integrating
out the bath degrees of freedom $Q_i$. The main assumption of this method is that the subsystem is uncorrelated with the bath at the initial time. Hence the total density matrix at time t = 0 is
given by $\rho_T(0) = \rho_A(0) \otimes \rho_B (0)$, where $\rho_A$, $\rho_B$ and $\rho_T$ are the
density matrix of the subsystem, bath and the total system respectively. However, several authors
have considered a correlated initial state, where the subsystem and bath density matrix cannot be
factorized. For example, Hakim and Ambegaokar \cite{hakim} have compared the two cases of 
uncorrelated and correlated initial conditions and shown that, for the correlated initial condition,
 different transient behaviours can be obtained at time scales larger than the inverse cutoff 
frequency of the bath, in contrast to the uncorrelated initial conditions. However, here we are
following the usual procedure of Feynman and Vernon and we will show finally that we can obtain a
consistent ME, describing the time evolution of the DM.
The time evolution of the reduced DM is given by, 
\begin{equation}
\rho(q_{1},q_{2},t)=\int dq_{1}^{\prime}dq_{2}^{\prime}J(q_{1},q_{2},t;q_{1}^{\prime},q_{2}^{\prime},0)\rho(q_{1}^{\prime},q_{2}^{\prime},0)
\label{time_evol}
\end{equation}
where the propagator $J(q_{1},q_{2},t;q_{1}^{\prime},q_{2}^{\prime},0)$ is given by, 
\begin{equation}
J(q_{1},q_{2},t;q_{1}^{\prime},q_{2}^{\prime},0)=\int
\int{\cal D}q_{1}{\cal D}q_{2} \exp\left(\frac{\imath}{\hbar} S_{\rm eff}[q_1,q_2]\right)
\label{propagator}
\end{equation}
After integrating out the bath degrees of freedom, one obtains the
nonlocal effective action corresponding to the dissipative system: 
\begin{equation}
\frac{\imath}{\hbar}S_{{\rm eff}}=\frac{\imath}{\hbar}
(S_{A}[q_{1}]-S_{A}[q_{2}])+\int_{0}^{t}(\Sigma_{R}+\Sigma_{I})d\tau
\end{equation}
where $S_A$ is the action corresponding to the non-interacting system, and,
\begin{eqnarray}
\Sigma_{R} & = & -\frac{1}{\hbar}\int_{0}^{\tau}[q_{-}(\tau)\alpha_{R}
(\tau-s)q_{-}(s)]ds\nonumber \\
\Sigma_{I} & = & -\frac{\imath}{\hbar}\int_{0}^{\tau}[q_{-}(\tau)\alpha_{I}(\tau-s)q_{+}(s)]ds\label{sigmas}
\end{eqnarray}
where, $q_{\pm}=q_{1}\pm q_{2}$ and the memory kernels are, 
\begin{eqnarray}
\alpha_{R}(\tau) & = & \sum_{i}\frac{C_{i}^{2}}{2m\omega_{i}}\coth
\left(\frac{\hbar\omega_{i}}{2k_{B}T}\right)\cos(\omega_{i}\tau)\\
\alpha_{I}(\tau) & = & -\sum_{i}\frac{C_{i}^{2}}{2m\omega_{i}}\sin(\omega_{i}\tau).
\nonumber
\label{memory-kernel} 
\end{eqnarray}

It is assumed that the oscillator frequencies are continuously distributed
from zero to a maximum frequency $\omega_{c}$, value of which depends
on specific physical system. Cutoff frequency of the heat bath is
chosen in such way that the characteristic time scale of the dynamics
of heavy particle is much larger than the collisional time scale $1/\omega_{c}$.
Thus, the summations in Eq. \ref{memory-kernel} can be replaced by integrals
by introducing an appropriate density of state, which will make the
memory function anaytically tractible. Hence, $\sum_{i}$ can be replaced y,
$\int d\omega F(\omega)$, where $F(\omega)$ is the density of states
of the bath. We choose a smooth Drude form for the density of states
given by $F(\omega)=\omega_{c}^{2}/(\omega^{2}+\omega_{c}^{2})$ and
$F(\omega)C^{2}(\omega)/(2m\omega^{2})=(2M\gamma/\pi)(\omega_{c}^{2}/(\omega^{2}+\omega_{c}^{2}))$.
At any non-zero temperature,using the Drude form of density of states the 
memory kernels can be evaluated
analytically by using contour integrals in complex $\omega$ plane and are given by:
\begin{eqnarray}
\alpha_{R}(\tau) & = & M\gamma \omega_{c}^{2}\left[\cot(\chi)\exp(-\omega_{c}\tau)\right.\nonumber \\
& + & \left.\frac{2}{\chi}\sum_{n=1}^{\infty}\frac{n\pi/\chi}{(n\pi/\chi)^{2}-1}\exp\left[-(n\pi/\chi)\omega_{c}\tau\right]\right]
\label{alpha-r-1}
\end{eqnarray}
\begin{equation}
\alpha_{I}(\tau)=M \gamma \omega_{c}\frac{\partial}{\partial \tau}\exp[-\omega_{c} \tau]
\label{alpha-i-1}
\end{equation}
where $\chi$ = $\hbar\omega_{c}/2k_{B}T$.
From the above expression of $\alpha_{R}$ we can clearly see the
emergence of two time scales - the first being the microscopic collision
time, given by $1/\omega_{c}$ and the other time scale given by temperature
i.e. $\hbar/k_{B}T$. It is interesting to note that in the high
temperature regime, $\alpha_{R}$ is short ranged over the thermal
time scale while the timescale over which $\alpha_{I}$ decays is
given only by the collision time scale, which is independent of temperature.
However, as the temperature is reduced, at some point, the thermal
time scale dominates over the collision time scale, giving rise to
a nonlocal memory kernel which has a power law decay.

We now proceed to simplify the nonlocal action, taking advantage of
the above mentioned short-ranged kernel. The dynamics over a timescale
larger than $\hbar/k_{B}T$ can be studied by assuming smooth classical
trajectories and expanding the dynamical variables in a Taylor series,
\begin{equation}
q_{\pm}(s)=\sum_{l=0}^{\infty}\frac{q_{\pm}^{(l)}(\tau)}{l!}(s - \tau)^{l}
\label{taylor}
\end{equation}
where, $q^{(l)}$ denotes the $l^{th}$ derivative of $q$ with respect
to time. We now insert Eq. \ref{alpha-r-1} and Eq. \ref{taylor},
into Eq. \ref{sigmas}. Neglecting the total derivative terms which
generate the boundary terms in the action, thereby leaving the Lagrangian
invariant, we obtain, 
\begin{equation}
\Sigma_{R}(\tau) = -\frac{1}{\hbar}\sum_{l=0}^{\infty}\frac{(-1)^{l}}{2l!} \left(q_{-}^{(l)}(\tau)\right)^2\int_{0}^{\tau}
\tilde \tau^{2l}\alpha_{R}(\tilde \tau)d\tilde \tau
\end{equation}
The transient term in the integral comes in the form of $\exp(-(k_{B}T/\hbar)\tau$.
Thus, for $\tau\gg\hbar/k_{B}T$, we obtain, 
\begin{eqnarray}
&&\Sigma_{R}(\tau) =  -\frac{M \gamma \omega_{c}}{\hbar\chi}\times \nonumber \\ 
 && \left[ 
q_{-}^{2}(\tau)
-2 \sum_{l=1}^{\infty}(q_{-}^{(l)})^{2}\frac{(-1)^{l}}{\omega_{c}^{2l}}
\sum_{m=0}^{l}\left(\frac{\chi}{\pi}\right)^{2m}\zeta(2m)\right] 
\label{sigma-r-gen}
\end{eqnarray}
Similarly, one can evaluate $\Sigma_{I}$ which gives, 
\begin{equation}
\Sigma_{I}(\tau)=-\frac{\imath\gamma M}{\hbar}q_{-}(\tau)\dot{q_{+}}(\tau)
\label{sigma-i-gen}
\end{equation}
where, the higher order terms in the expansion of $q_{+}(s)$ have
been neglected, since they fall off as $1/\omega_{c}$.

We have thus transformed a highly nonlocal action into a local action
containing higher derivative terms. While in general, this may seem
to be a very complicated expression to work with, we show below that
in at least the case of a particle in a harmonic potential attached
to a heat bath and a Brownian particle there is considerable simplification
of the expression which allows us to obtain exact results.


For the free particle, general dynamical equation for any order of time derivative 
of the position coordinate can be written as, 
\begin{equation}
q^{(n)}=\frac{(-\gamma)^{n-1}}{M}p
\label{br-cl-eq-mot}
\end{equation}
Substituting Eq. \ref{br-cl-eq-mot} in Eq. \ref{sigma-r-gen}, we obtain,
\begin{eqnarray}
&&\Sigma_{R}(\tau)  =  -\frac{M\gamma\omega_{c}}{\hbar\chi} q_{-}^{2}(\tau) \nonumber \\ 
&-& \frac{M\gamma\omega_{c}}{\hbar\chi} \left[ \frac{\dot{q_{-}}^{2}(\tau)}{\gamma^{2}}\left\{\frac{(\frac{\gamma
\chi}{\omega_{c}})\coth\left(\frac{\gamma\chi}{\omega_{c}}\right)}{1+\left(\frac{\gamma}{\omega_{c}}\right)^{2}}
-1\right\}\right]
\end{eqnarray}
Thus, the effective action is given by, 
\begin{eqnarray}
\frac{\imath}{\hbar}S_{{\rm eff}}  &=& 
 \frac{\imath}{\hbar}\int_{0}^{t} d\tau \left[\frac{M}{2}\dot{q}_{+}
\dot{q}_{-} -\gamma M q_{-} \dot{q}_{+} \right]\nonumber\\
&-&\frac{2 k_{B} T \gamma M}{\hbar^{2}} \int_{0}^{t} d\tau \left[ q_{-}^{2}
+ \alpha \dot{q}_{-}^{2} \right]
\end{eqnarray}
where $\alpha = [\frac{\hbar \gamma}{2 k_{B} T} \coth(\frac{\hbar \gamma}
{2 k_{B} T}) - 1]/\gamma^{2}$, assuming $\gamma \ll \omega_{c}$.
It is important to note that by using the dynamical equation of canonical
coordinates and resummation method we have converted the non-local action into an 
effective action which is local and quadratic in canonical coordinates as well as
independent of the cut-off frequency $\omega_{c}$ of bath oscillators. Inserting the 
above form of the effective action into Eq. \ref{propagator} the time evolution of the 
DM can be evaluated.
%

Following the prescription of
CL \cite{cl}, we consider the change of density matrix from $t$ to $t + \epsilon$ within a small time interval $\epsilon$,
in order to obtain the ME in
differential form. 
To do so we expand both side of Eq. \ref{time_evol} upto leading
order in $\epsilon$.
Within the small time $\epsilon$, we approximate $\dot{q}_{1} = (x - x')/\epsilon=\beta_{1}/\epsilon$ and $\dot{q}_{2} = (y-y')/\epsilon = \beta_{2}/\epsilon$.
Now Eq. \ref{time_evol} reads:
\begin{eqnarray}
\rho + \epsilon \frac{\partial \rho}{\partial t} = \int\int d\beta_{+}
d \beta_{-} exp\left[\frac{i}{\hbar \epsilon}M\beta_{+}(\beta_{-} 
- 2\gamma x_{-} \epsilon) \right.\nonumber\\
\left. -\frac{2 k_{B} T \gamma M}{\hbar^{2}}(x_{-}^{2}
+ \alpha \beta_{-}^{2})\right] \rho(x-\beta_{1},y-\beta_{2},t),
\label{rho_exp}
\end{eqnarray}
where $\beta_{+} = (\beta_{1} + \beta_{2})/2$ and $\beta_{-} = \beta_{1} - 
\beta_{2}$.
We expand $\rho$ upto second order in  $\beta$, which is equivalent to
expanding upto first order in $\epsilon$.
 After doing some algebra and performing the gaussian integrals
we obtain the ME
describing the time evolution of the density matrix:
\begin{eqnarray}
\frac{\partial \rho}{\partial t} &=& \frac{i \hbar}{2 M}\left[ \frac{\partial^{2}
\rho}{\partial x^2} -  \frac{\partial^{2}
\rho}{\partial y^2} \right] -\gamma (x -y)(\frac{\partial \rho}{\partial x} -
\frac{\partial \rho}{\partial y})\nonumber \\
&+& \frac{2\imath}{\hbar}D_{pq}(x -y)
(\frac{\partial \rho}{\partial x} 
+ \frac{\partial \rho}{\partial y}) +
D_{qq}(\frac{\partial}{\partial x} +  
\frac{\partial}{\partial y})^{2}\rho \nonumber \\ 
&-& \frac{D_{pp}}{\hbar^2}(x - y)^{2} \rho.
\end{eqnarray}
where, different diffusion constants are given by:
\begin{eqnarray}
D_{qq} & = & \frac{2k_BT}{M\gamma}\left[ \left(\frac{\hbar \gamma}{2 k_B T}\right)\coth\left(\frac{\hbar \gamma}{2 k_B T}\right) -1 \right] \\
D_{pq} & = & 4 k_BT \left[ \left(\frac{\hbar \gamma}{2 k_B T}\right)\coth\left(\frac{\hbar \gamma}{2 k_B T}\right) -1 \right]\\
D_{pp} & = & 2M\gamma k_B T \left[ \left(2\frac{\hbar \gamma}{k_B T}\right)\coth\left(\frac{\hbar \gamma}{2 k_B T}\right) -3\right]
\end{eqnarray}
The above form of the ME has been proposed earlier phenomenologically 
\cite{dekker} and the diffusion constants have to satisfy some conditions, 
in order to preserve the 
Heisenberg Uncertainty Principle in quantum dissipative systems
\cite{positivity1}. 
This form of the ME can also be recast into the Lindblad form 
\cite{positivity2} which maintains
the positivity of DM during time evolution. Commonly known 
{\it positivity condition} is 
described by\cite{positivity1,positivity2} :
\begin{equation}
\Delta = D_{pp} D_{qq} - D_{pq}^2 - \frac{\hbar^2 \gamma^2}{4} \ge 0
\end{equation}
At high temperatures, the diffusion constants behave as,
$D_{pp} = 2M\gamma k_B T \left[1+\frac{2}{3}\left(\frac{\hbar \gamma}{k_B T}\right)^2\right]$,
$D_{qq} = \frac{\hbar^2 \gamma}{6 M k_B T}$ and 
$D_{pq} = \frac{\hbar^2\gamma^2}{3 k_B T}$.
The forms of these diffusion constants are very similar to those obtained 
earlier \cite{diosi}. However, we would like to point out that the diffusion constant $D_{pq}$, in our case, is independent of the cutoff parameter $\omega_c$, unlike the earlier result. It is also interesting to note that the diffusion constants $D_{qq}$ and $D_{pp}$ both originate from quantum effects and vanish at 
high temperatures as $1/k_B T$. However, at high temperatures, $\Delta$ 
approaches a value $\hbar^2 \gamma^2/12$ which preserves the positivity 
condition of the ME. At zero temperature, the diffusion 
constants are finite, positive and proportional to $\hbar$, 
similar to the form proposed by Dekker \cite{dekker}. 
However, $\Delta$ becomes negative at zero temperature, violating the 
positivity criterion. We estimate a critical temperature $T_{0} \sim
0.2 \hbar \gamma$, below which above form of ME is not valid and transient
behaviour as well as long range memory effects become crucial.

To summarise, within our scheme of calculation, we have shown that the ME 
obtained from the microscopic CL model satisfies the positivity condition and belong to the Lindblad class above a certain temperature. Analytical form for the diffusion constants have been obtained for any arbitrary temperature and 
are {\it independent of the cutoff frequency of the heat bath}.
It is interesting to note that at high temperatures all the anomalous 
diffusion constants vanish as
$1/T$, which preserves the structure of classical Brownian motion. At the 
same time, the diffusion constants conspire in such a way that they satisfy 
the Dekker criterion $\Delta > 0$, hence the time evolution of the open 
quantum system takes place only through physical states. At low temperature, 
the diffusion constants have their origin from purely quantum effects. 
However, below the temperature $\simeq \hbar \gamma$, Dekker's positivity 
condition is violated, indicating that the long range memory effect may 
become important in the time evolution of the system.

We like to thank P. K. Panigrahi and S. Dattagupta for helpful discussions
and critical reading of the manuscript.

\end{document}